\documentclass[letterpaper]{article} 
\usepackage{aaai25}  
\usepackage{times}  
\usepackage{helvet}  
\usepackage{courier}  
\usepackage[hyphens]{url}  
\usepackage{graphicx} 
\usepackage{natbib}  
\usepackage{caption} 
\usepackage{algorithm}
\usepackage{algorithmic}

\usepackage{multirow}
\usepackage{colortbl}
\usepackage{xcolor}
\usepackage{amsmath}
\usepackage{amssymb}

\usepackage{newfloat}
\usepackage{listings}
\DeclareCaptionStyle{ruled}{labelfont=normalfont,labelsep=colon,strut=off} 
\floatstyle{ruled}
\newfloat{listing}{tb}{lst}{}
\floatname{listing}{Listing}
\newif\ifarxiv
\arxivtrue

\ifarxiv
\usepackage{fancyhdr}
\fi

\title{Wavelet-Assisted Multi-Frequency Attention Network for Pansharpening}

\author{
    Jie Huang\textsuperscript{\rm 1}\thanks{Co-first authors contributed equally.},
    Rui Huang\textsuperscript{\rm 1}\footnotemark[1],
    Jinghao Xu\textsuperscript{\rm 1},
    Siran Peng\textsuperscript{\rm 2, \rm 3},
    Yule Duan\textsuperscript{\rm 1},
    Liangjian Deng\textsuperscript{\rm 1}\thanks{ Corresponding author.}
}

\affiliations{
    \textsuperscript{\rm 1}University of Electronic Science and Technology of China\\
    \textsuperscript{\rm 2}Institute of Automation, Chinese Academy of Sciences\\
    \textsuperscript{\rm 3}School of Artificial Intelligence, University of Chinese Academy of Sciences\\
    \{jayhuang,huang\_rui,2023040901014\}@std.uestc.edu.cn,pengsiran2023@ia.ac.cn, yule.duan@outlook.com,liangjian.deng@uestc.edu.cn
}

\usepackage{fancyhdr}
\begin{document}

\maketitle

\begin{abstract}
Pansharpening aims to combine a high-resolution panchromatic (PAN) image with a low-resolution multispectral (LRMS) image to produce a high-resolution multispectral (HRMS) image. Although pansharpening in the frequency domain offers clear advantages, most existing methods either continue to operate solely in the spatial domain or fail to fully exploit the benefits of the frequency domain. To address this issue, we innovatively propose Multi-Frequency Fusion Attention (MFFA), which leverages wavelet transforms to cleanly separate frequencies and enable lossless reconstruction across different frequency domains. Then, we generate Frequency-Query, Spatial-Key, and Fusion-Value based on the physical meanings represented by different features, which enables a more effective capture of specific information in the frequency domain. Additionally, we focus on the preservation of frequency features across different operations. On a broader level, our network employs a wavelet pyramid to progressively fuse information across multiple scales. Compared to previous frequency domain approaches, our network better prevents confusion and loss of different frequency features during the fusion process. Quantitative and qualitative experiments on multiple datasets demonstrate that our method outperforms existing approaches and shows significant generalization capabilities for real-world scenarios. 

\end{abstract}


\begin{links}
    \link{Code}{https://github.com/Jie-1203/WFANet}
\end{links}

\section{Introduction}
High-resolution multispectral (HRMS) images are vital for applications like environmental monitoring and urban planning. Due to hardware constraints, satellites typically capture low-resolution multispectral (LRMS) and high-resolution panchromatic (PAN) images. Pansharpening fuses these to produce HRMS, combining their strengths to enhance spatial and spectral resolution.

To obtain high-resolution multispectral (HRMS) images, pansharpening methods are categorized into traditional and deep learning-based approaches. Traditional methods are divided into three groups \cite{overall}: Component Substitution (CS) \cite{CS},
\begin{figure}[!h]
\centering
\includegraphics[width=\columnwidth]{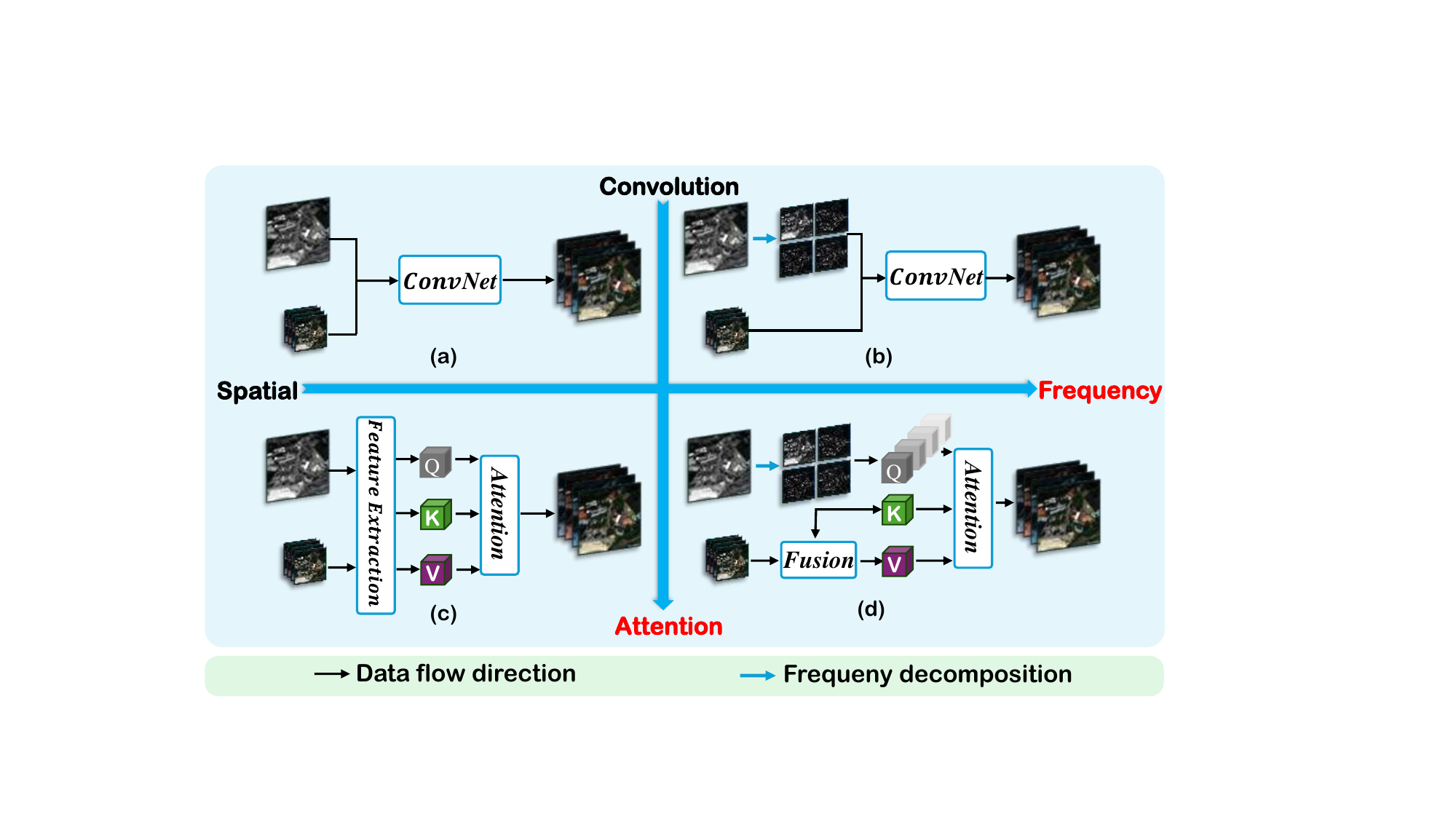} 
\caption{The comparison covers four methods across two dimensions: (a) Convolutional network in the spatial domain,
(b) Convolutional network in different frequency domains,
(c) Attention mechanism in the spatial domain, 
and (d) Our proposed method which forms the primary motivation for this paper: 1) utilizing wavelet transforms to process in different frequency domains; 2) designing an attention method with clear physical significance to leverage the advantages of frequency domain processing.}
\label{your_label}
\end{figure}
 Multi-Resolution Analysis (MRA) \cite{MRA}, and Variational Optimization-based (VO) \cite{VO} techniques. In recent years, with the rapid development of deep learning, many deep learning methods \cite{SSconv, Triple} have been proposed for pansharpening using convolutional neural networks (CNN), as shown in Fig.~\ref{your_label} (a), such as PNN \cite{PNN}, DiCNN \cite{DiCNN}, and LAGConv \cite{LAGConv}. These methods underscore deep learning's potential to improve pansharpening accuracy and efficiency. However, most existing methods do not process images in different frequency domains but instead operate in the original single spatial domain, thereby limiting the potential for improving fusion quality.

\begin{figure}[t]
\centering
\includegraphics[width=0.99\columnwidth]{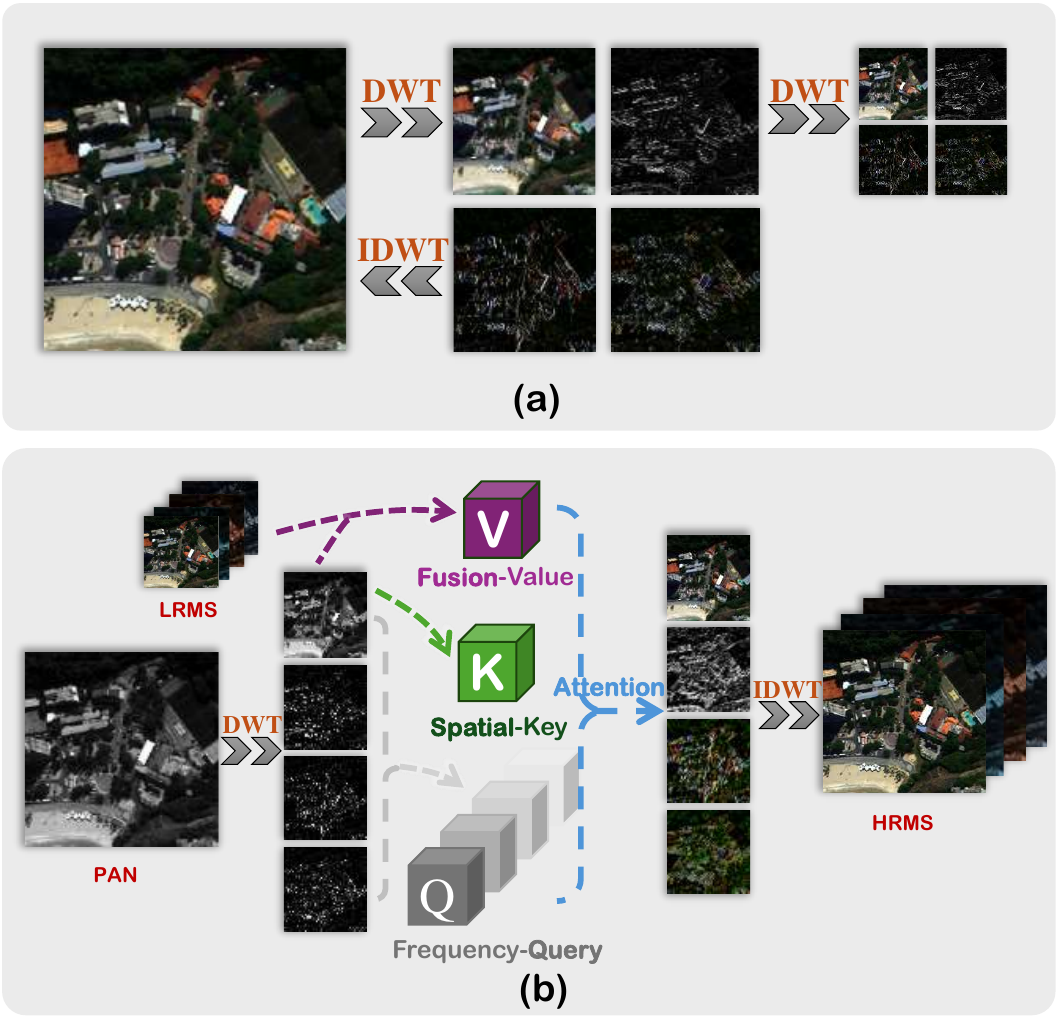} 
\caption{
(a) DWT decomposes the image into four different frequency components. IDWT is the lossless inverse process of DWT. Multiple applications of DWT produce a multi-scale wavelet pyramid. 
(b) Simplified illustration of MFFA. Fusion-Value, Spatial-Key, and Frequency-Query are derived from the information indicated by the arrows. These components are then processed through an attention mechanism, enabling the reconstruction of features across different frequencies that integrate both spectral and spatial information.}
\label{fig1}
\end{figure}

Direct fusion in the spatial domain methods can often result in detail loss or blurring due to the imprecise separation of frequency information. 
In contrast, frequency-based methods can separate different frequencies for targeted processing, which better preserves hard-to-capture high-frequency information while effectively preventing interference between different frequencies.
Consequently, processing in different frequency domains can be a more effective approach for achieving better fusion results compared to spatial domain fusion methods, and some works \cite{jin2022laplacian, GuidedNet} have already attempted this approach. 
For example, AFM-DIN \cite{detail-injection} introduces a high-frequency injection module to enhance LRMS features with PAN details, but it may lose low-frequency information. FAMENet \cite{MOE} uses an expert mixture model to fuse different frequencies, effectively balancing both high-frequency and low-frequency information.
However, these methods often struggle to achieve clean separation, leading to interference and information loss due to the neural network's tendency to slightly blend frequency components together \cite{shan2021decouple}. To address these issues, we adopt a method that cleanly separates frequencies and enables lossless reconstruction.
As shown in Fig.~\ref{fig1} (a), the Discrete Wavelet Transform (DWT) cleanly separates frequency components \cite{wave1,wave2}. The Inverse Discrete Wavelet Transform (IDWT) is lossless, preserving all information. Repeated DWT builds a wavelet pyramid \cite{wave3}, efficiently handling multi-scale features and enhancing detail detection, offering advantages over other methods that attempt to separate 
frequency components.

Using wavelet transforms to fuse information across different frequency domains is an innovative approach. Designing appropriate networks and modules to effectively extract and combine these features is therefore crucial for achieving optimal fusion results. Some past methods have employed wavelet transforms for pansharpening, as shown in Fig.~\ref{your_label} (b), 
such as FAFNet \cite{FAFNet}, which utilizes DWT layers to extract and manage frequency-domain features, followed by IDWT layers that reconstruct these features into the spatial domain, with the final fusion achieved through a convolutional block that produces the high-quality fused image.
However, convolutional neural networks inherently excel at capturing low-frequency information and are less effective in capturing high-frequency details \cite{xu2019,CVIP}. This limitation leads to decreased fusion quality. 
To address this issue, we consider leveraging attention mechanisms that can more flexibly capture specific features \cite{soydaner2022attention}.
Some previous methods \cite{hou2024linearly,bidirectional} have attempted to use attention mechanisms in the spatial domain, as shown in Fig.~\ref{your_label} (c).
For example, PanFormer \cite{panformer} employs a customized Transformer architecture, using panchromatic (PAN) and low-resolution multispectral (LRMS) features as queries and keys for joint feature learning, modeling long-range dependencies to produce high-quality pan-sharpened results. 
However, such a design does not explicitly capture and fuse information from different frequencies.
In other visual domains, combining wavelet transforms with attention mechanisms, such as Wave-ViT \cite{wave-vit}, integrates wavelet transforms with Transformers. By performing invertible down-sampling within Transformer blocks, this method improves image recognition and enhances visual representation accuracy.

\begin{figure*}[!h]
\centering
\includegraphics[width=0.99\textwidth]{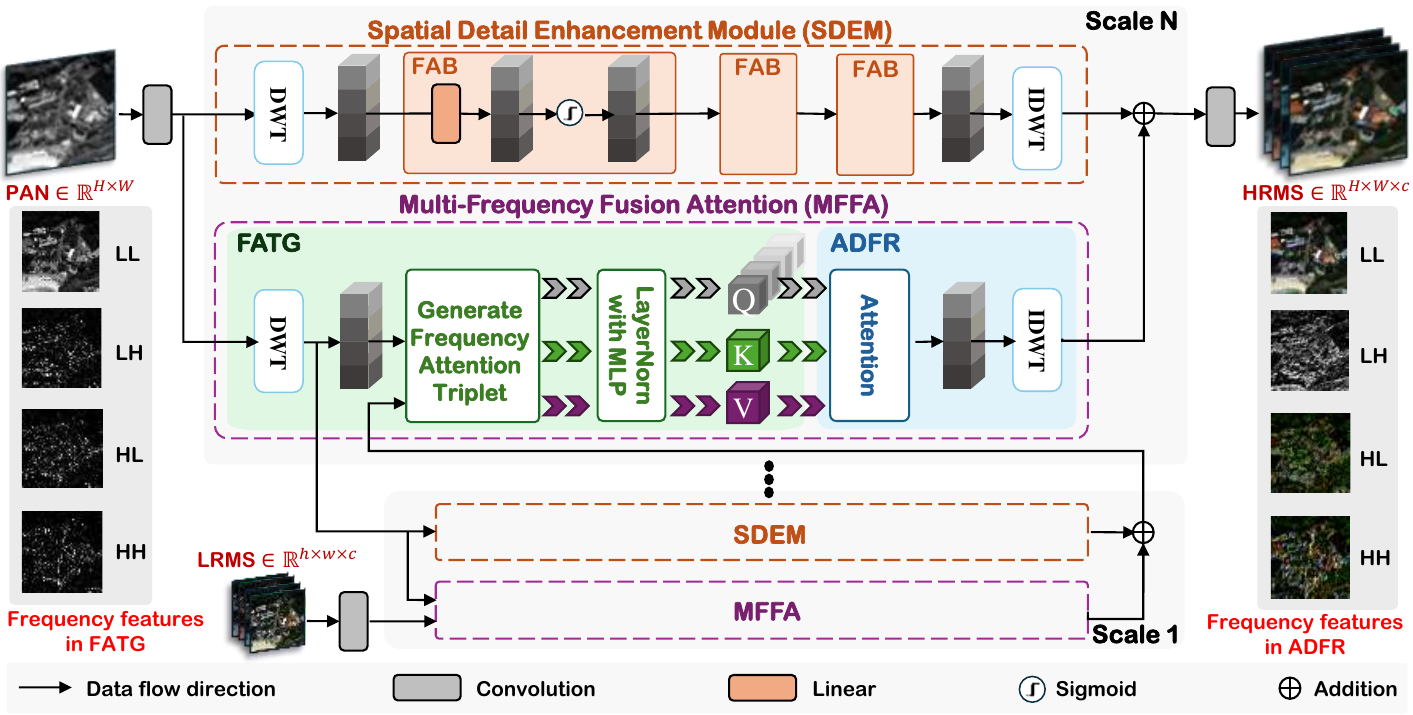} 
\caption{The overall workflow of our WFANet. Our network processes the data using multiple scales (only two scales are illustrated here for simplicity). WFANet consists of two sub-modules: the Multi-Frequency Fusion Attention (MFFA) and the Spatial Detail Enhancement Module (SDEM). The illustration of frequency features is shown on both sides of the figure.} 
\label{fig3}
\end{figure*}

Inspired by the above discussion, we design an innovative Wavelet-Assisted Multi-Frequency Attention Network called WFANet, \textit{with the core component being the Multi-Frequency Fusion Attention (MFFA).}
In our proposed MFFA, \textit{we introduce the concept of a Frequency Attention Triplet, which consists of Frequency-Query, Spatial-Key, and Fusion-Value}. As shown in Fig.~\ref{fig1} (b), we apply DWT to the PAN image, achieving a clean separation of different frequency features, effectively avoiding interference between them. To process and fuse this separated frequency information, unlike conventional attention mechanisms \cite{vaswani2017attention,soydaner2022attention}, our Frequency Attention Triplet has distinct physical significance: Frequency-Query represents frequency features, Spatial-Key encodes spatial information, and Fusion-Value represents the preliminary fusion of spatial and spectral features. This design effectively guides the information fusion process. The attention mechanism then captures correlations to achieve the initial fusion of frequency features, and IDWT is used for lossless reconstruction. Through MFFA, we can more effectively fuse information across different frequency domains and prevent the loss of frequency.
Additionally, inspired by previous methods \cite{FUsionNet, hou2023bidomain, Variational}, which demonstrated that enhancing spatial details significantly improves restoration when a module is primarily focused on fusion, as realized in our MFFA, \textit{we design the Spatial Detail Enhancement Module (SDEM) to focus on the extraction and enhancement of spatial details}.
In designing SDEM, we compare different operations for their adaptability to the frequency domain, ensuring better preservation of spatial details.
Moreover, our overall framework is a multi-scale progressive reconstruction framework, fully utilizing the inherent multi-scale nature of the wavelet pyramid.

In summary, the contributions of this work are as follows:
\begin{itemize}
    \item  
We introduce the Multi-Frequency Fusion Attention (MFFA), utilizing wavelet transforms to cleanly separate and accurately reconstruct frequency components. This approach integrates the Frequency-Query, Spatial-Key, and Fusion-Value triplet to enhance feature fusion precision across different frequency domains, effectively reducing confusion and information loss.
    \item Additionally, we focus on how different operations preserve frequency features and utilize the wavelet pyramid for progressive, multi-scale fusion. The effectiveness of these strategies has been validated and demonstrated through extensive ablation experiments.
    \item Our method achieves state-of-the-art performance on three diverse pansharpening datasets, demonstrating high-quality fusion results supported by both quantitative and qualitative experimental evidence.
\end{itemize}

\begin{figure*}[!h]
\centering
\includegraphics[width=0.99\textwidth]{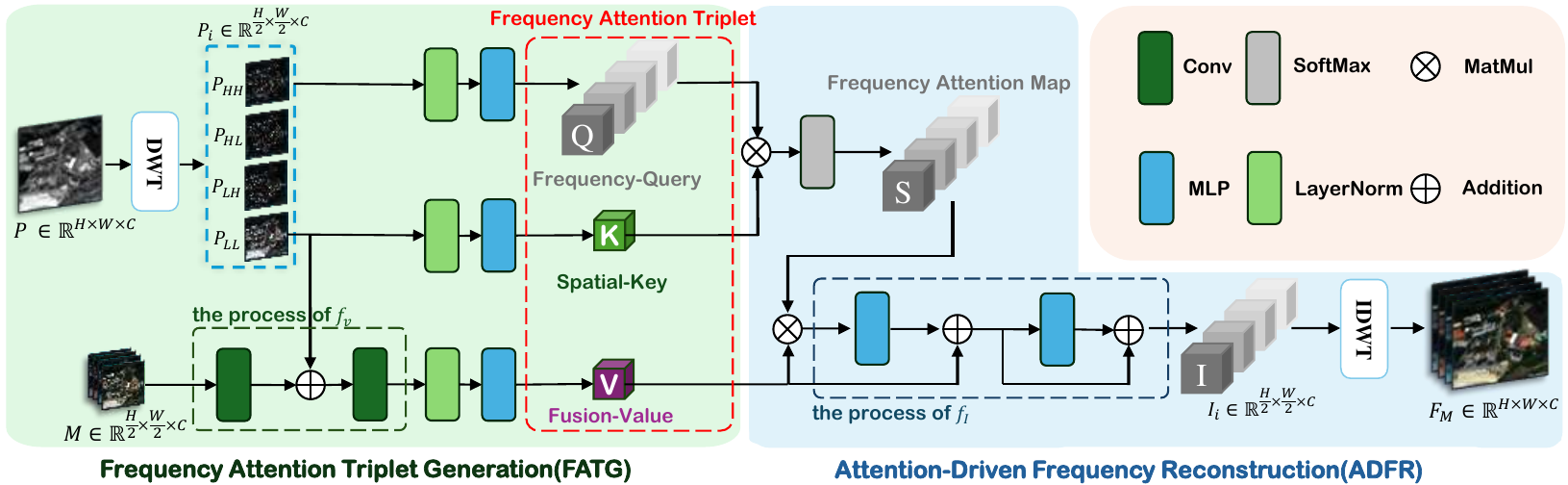} 
\caption{
The MFFA workflow involves two phases. First, in the FATG phase, the Frequency Attention Triplet with specific physical significance is generated. Then, in the ADFR phase, the obtained Frequency Attention Triplet is processed to reconstruct the features at different frequencies. \( Q \), \( S \), and \( I \) are shown as four colored blocks, representing features from different frequency domains after DWT. The data dimensions are exemplified using the largest scale.}
\label{fig4}
\end{figure*}

\section{Proposed Method}
This section introduces the proposed WFANet, detailing its two key components: Multi-Frequency Fusion Attention (MFFA) and the Spatial Detail Enhancement Module (SDEM), followed by the overall multi-scale framework. Fig.~\ref{fig3} illustrates the workflow of WFANet.

\subsection{Multi-Frequency Fusion Attention (MFFA)}
To fuse information across frequencies, we propose the MFFA, which is composed of two phases: Frequency Attention Triplet Generation (FATG) and Attention-Driven Frequency Reconstruction (ADFR). Details are shown in Fig.~\ref{fig4}.

\subsubsection{(I) Frequency Attention Triplet Generation}

As in the typical attention mechanism \cite{vaswani2017attention,soydaner2022attention}
, Query, Key, and Value are the three components of attention. Query represents the information we seek, Key is the index of this information, and Value is the specific content. To better adapt to different frequency domains, we design the Frequency Attention Triplet. Specifically, different frequency features are the information we seek, the overall spatial features are the indices for querying different frequency features, and the specific content is the fusion of spectral and spatial information. Therefore, we design Frequency-Query, Spatial-Key, and Fusion-Value with specific physical meanings. We first perform a DWT on \( P \), which is the feature of the panchromatic (PAN) image after convolution, as shown below:

\begin{equation}
P_{LL}, P_{LH}, P_{HL}, P_{HH} = \operatorname{DWT}(P)
\label{DWT}
\end{equation}
where \( P_{LL} \) represents the low-frequency details of \( P \), and \( P_{LH} \), \( P_{HL} \), and \( P_{HH} \) represent the high-frequency details of \( P \) in the horizontal, vertical, and diagonal directions, respectively.
\textit{For convenience, \( i = LL, LH, HL, HH \) corresponds to the four frequency features mentioned above, respectively.}
The DWT operation has already separated the frequency features adequately, 
thus we directly use \( P_i \), representing different frequency features, as \( Q_i \).
Low-frequency features represent the overall spatial appearance of the image, 
while high-frequency information represents edge details and fine textures \cite{citation1}.
Therefore, to use the overall spatial features as our key, 
we directly take the low-frequency features represented by \( P_{LL} \) as the Spatial-Key.
Then, we design an \( f_v \) to fuse spatial and spectral information as the Fusion-Value.
Next, the three components are normalized using LayerNorm and processed through an MLP to enhance their expressiveness.
The specific process can be described by the following equations:
\begin{equation}
\begin{aligned}
    &Q_i = \operatorname{MLP}(\operatorname{LN}(P_i)) \\
    &K = \operatorname{MLP}(\operatorname{LN}(P_{LL})) \\
    &V = \operatorname{MLP}(\operatorname{LN}(f_v(M, P_{LL})))
\end{aligned}
\end{equation}
where  \( M \) represents the feature of the low-resolution multispectral (LRMS) image after convolution and \( f_v \) represents the process of obtaining the Fusion-Value by combining \( M \) and \( P_{LL} \) through convolution.

\subsubsection{(II) Attention-Driven Frequency Reconstruction}
After obtaining the Frequency Attention Triplet, we use the attention mechanism to reconstruct the fused features at different frequencies.
First, calculate the frequency correlation \( R_i \) between the Frequency-Query and Spatial-Key, representing the correlation between different frequency features and the overall spatial features. Then, apply softmax to obtain the Frequency Attention Map \( S_i \), which highlights the importance of different frequency features relative to the overall spatial features. The process is as follows:
\begin{equation}
\begin{aligned}
R_i &= Q_i \otimes K \\
S_i &= \operatorname{softmax}(R_i)
\end{aligned}
\end{equation}
where  \(\otimes\) represents matrix multiplication and softmax is an operation that converts input values into a probability distribution.
Next, the Frequency Attention Map \( S_i \) of different frequencies is separately multiplied by \( V \), which is the fusion of spectral information and low-frequency spatial information. Then, the result is processed through MLPs and residual connection to obtain the reconstructed features of different frequencies containing spectral information \( I_i \). This process can be described by the following equation:

\begin{figure}[t]
\centering
\includegraphics[width=0.75\columnwidth]{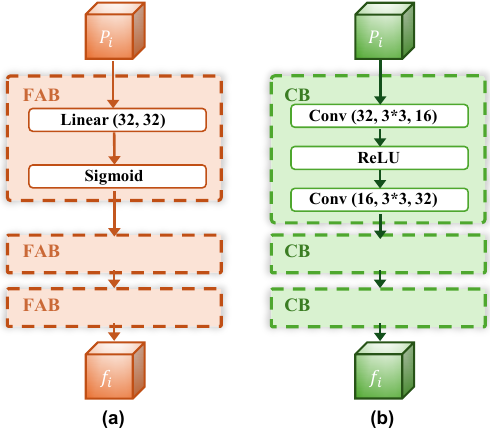} 
\caption{Comparison of two network architectures for the SDEM:
(a) Frequency Adaptation Block (FAB), which is used in the SDEM.
(b) Convolution Block (CB).
} 
\label{comparison}
\end{figure}

\begin{equation}
I_i = f_I(S_i \otimes V)
\end{equation}
where \( f_I \) represents the process of applying MLPs and residual connection to the result of the multiplication.
Finally, the preliminary reconstructed image \( F_M \) is obtained by leveraging the lossless property of the IDWT, as shown below:

\begin{equation}
F_M = \operatorname{IDWT}(I_{LL}, I_{LH}, I_{HL}, I_{HH})
\end{equation}
where \( I_{LL} \), \( I_{LH} \), \( I_{HL} \), and \( I_{HH} \) represent the features reconstructed at different frequencies.

\subsection{Spatial Detail Enhancement Module (SDEM)}

The core component, MFFA, achieves the fusion of information across different frequency domains. 
In contrast, SDEM focuses on extracting and enhancing spatial detail information within these frequency domains.
First, we decompose \( P \) according to Eq.\ref{DWT} to obtain \( P_i \), features containing information from different frequencies, helping to prevent interference between them. Next, we extract \( f_i \), representing spatial information for different frequency features, separately using several Frequency Adaptation Blocks (FABs). An FAB is a block capable of adapting to different frequencies and is composed of a linear layer and a sigmoid activation function. This process is illustrated below:

\begin{equation}
\begin{aligned}
    f_i &= \operatorname{FABs}(P_i)
\end{aligned}
\end{equation}
where \( i = LL, LH, HL, HH \) correspond to the four different frequency features, respectively. As illustrated in Fig.~\ref{comparison}, we do not choose the Convolution Block. Given that convolutional networks struggle with extracting high-frequency information and perform poorly when processing different frequency domains \cite{xu2019,CVIP}, we opt for linear layers, which, as demonstrated by our ablation experiments, better adapt to different frequency domains. After extracting features in different frequency domains, we use the lossless IDWT to recover the complete spatial details \( F_S \), as illustrated below:

\begin{equation}
F_{S} = \operatorname{IDWT}(f_{LL}, f_{LH}, f_{HL}, f_{HH})
\end{equation}

\subsection{Network Framework and Loss}

This section describes how to utilize the wavelet pyramid to 
construct the multi-scale network architecture of WFANet with MFFA and SDEM.
Our network employs a multi-scale structure with \( N \) layers (limited by the dataset, we use two scales in this paper).
First, we construct a wavelet pyramid by repeatedly applying DWT, as follows:
\begin{equation}
P_{k} = \operatorname{DWT}(P_{k+1})
\label{down}
\end{equation}
where \( P_{k+1} \) represents the four different frequency features of the previous larger scale, and \( P_k \) represents the frequency features of the next smaller scale.
Then, we progressively fuse from the smallest scale.
\( P_k \) and \( M_k \) are the inputs at the \( k \)-th scale.
The output of each layer is obtained by adding the output \( F_M \) from the MFFA and the output \( F_S \) from the SDEM.
The output of the \( k \)-th layer serves as the input \( M_{k+1} \) for the next layer, which in turn enables the process of progressive reconstruction, expressed as follows:

\begin{equation}
\left\{
\begin{aligned}
    &M_1 = f(M_0, P_0) \\
    &M_2 = f(M_1, P_1) \\
    &\phantom{M_1, P_1} \vdots \\
    &M_n = f(M_{n-1}, P_{n-1}) 
\end{aligned}
\right.
\end{equation}
where \(n\) represents the number of scales. \(M_n\) is then convolved to get the final high-resolution multispectral (HRMS) image \(\hat{M}\). 
We choose the simple $\ell_{1}$ loss function since it is sufficient to yield consistently good outcomes: 
\begin{equation}
\label{loss}
\mathcal{L} = \frac{1}{K}\sum_{i=1}^{K} \| \hat{M}^{\{i\}} - I^{\{i\}} \|_1 
\end{equation}
where \(K\) is the number of training data, \( I^{\{i\}} \) denotes the $i$-th ground truth image, and \( \|\cdot\|_1 \) represents the $\ell_{1}$ norm.

\section{Experiments}

\begin{figure*}[!h]
\centering
\includegraphics[width=0.99\textwidth]{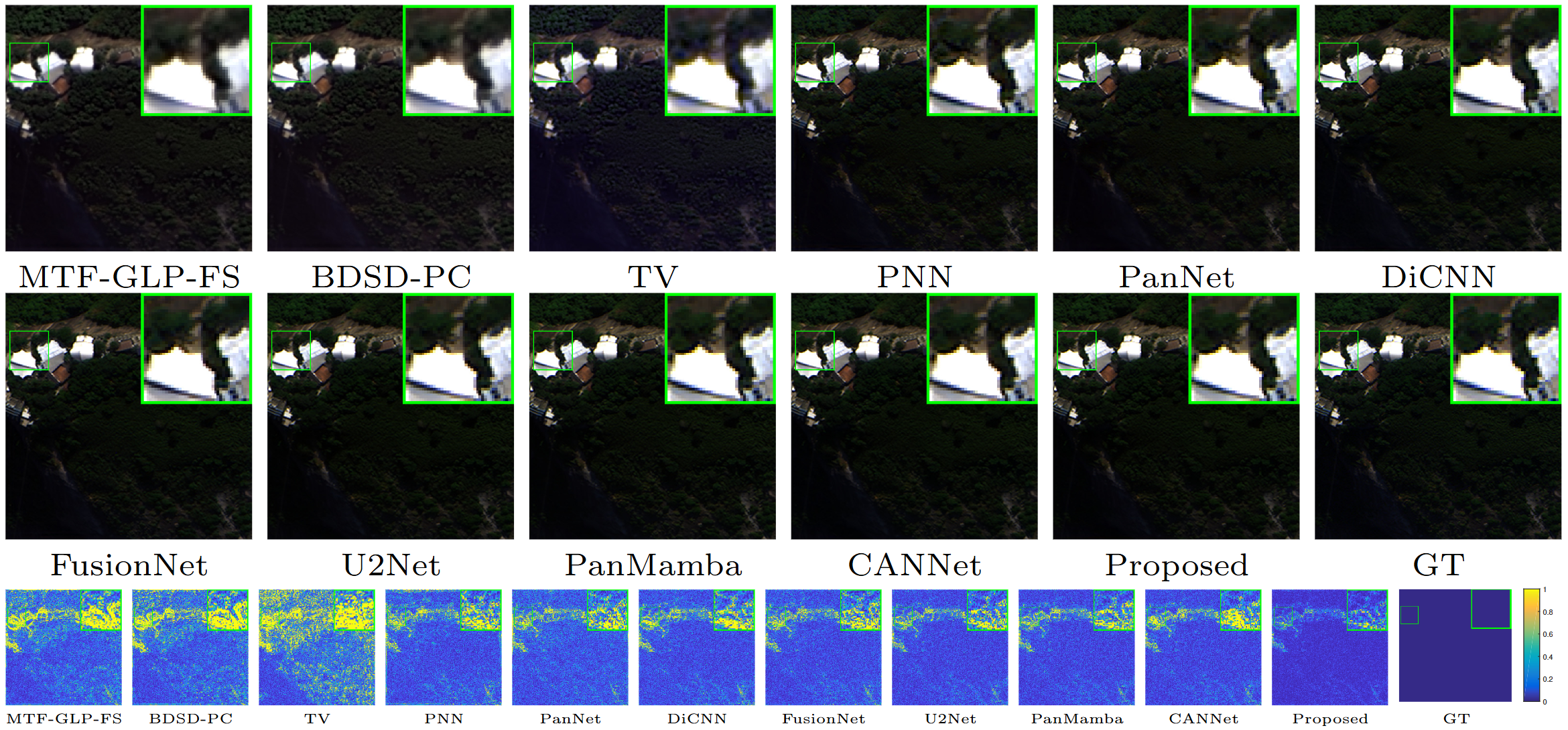} 
\caption{ The visual results (Top) and residuals (Bottom) of all compared approaches on the WV3 reduced-resolution dataset.} 
\label{qualitative-wv3}
\end{figure*}

\begin{table*}[h]
\centering
\begin{tabular}{c@{\hskip 0.02in}c@{\hskip 0.02in}c@{\hskip 0.02in}c@{\hskip 0.02in}c|c@{\hskip 0.02in}c@{\hskip 0.02in}c@{\hskip 0.02in}c|c@{\hskip 0.02in}c@{\hskip 0.02in}c@{\hskip 0.02in}c}
\hline
\multirow{2}{*}{\textbf{Methods}} & \multicolumn{4}{c|}{\textbf{WV3 }} & \multicolumn{4}{c|}{\textbf{QB }} & \multicolumn{4}{c}{\textbf{GF2 }} \\  
 & \textbf{PSNR$\uparrow$} & \textbf{SAM$\downarrow$} & \textbf{ERGAS$\downarrow$} & \textbf{Q8$\uparrow$} & \textbf{PSNR$\uparrow$} & \textbf{SAM$\downarrow$} & \textbf{ERGAS$\downarrow$} & \textbf{Q4$\uparrow$} & \textbf{PSNR$\uparrow$} & \textbf{SAM$\downarrow$} & \textbf{ERGAS$\downarrow$} & \textbf{Q4$\uparrow$} \\ \hline
MTF-GLP-FS & 32.963 & 5.316 & 4.700 & 0.833 & 32.709 & 7.792 & 7.373 & 0.835 & 35.540 & 1.655 & 1.589 & 0.897 \\  
BDSD-PC & 32.970 & 5.428 & 4.697 & 0.829 & 32.550 & 8.085 & 7.513 & 0.831 & 35.180 & 1.681 & 1.667 & 0.892 \\ 
TV & 32.381 & 5.692 & 4.855 & 0.795 & 32.136 & 7.510 & 7.690 & 0.821 & 35.237 & 1.911 & 1.737 & 0.907 \\ \hline

PNN & 37.313 & 3.677 & 2.681 & 0.893 & 36.942 & 5.181 & 4.468 & 0.918 & 39.071 & 1.048 & 1.057 & 0.960 \\ 
PanNet & 37.346 & 3.613 & 2.664 & 0.891 & 34.678 & 5.767 & 5.859 & 0.885 & 40.243 & 0.997 & 0.919 & 0.967 \\ 
DiCNN & 37.390 & 3.592 & 2.672 & 0.900 & 35.781 & 5.367 & 5.133 & 0.904 & 38.906 & 1.053 & 1.081 & 0.959 \\ 
FusionNet & 38.047 & 3.324 & 2.465 & 0.904 & 37.540 & 4.904 & 4.156 & 0.925 & 39.639 & 0.974 & 0.988 & 0.964 \\

U2Net & \underline{39.117} & \underline{2.888} & \underline{2.149} & \underline{0.920} & 38.065 & 4.642 & 3.987 & 0.931 & 43.379 & 0.714 & 0.632 & 0.981 \\ 
PanMamba & 39.012 & 2.913 & 2.184 & 0.920 & 37.356 & 4.625 & 4.277 & 0.929 & 42.907 & 0.743 & 0.684 & 0.982 \\ 
CANNet & 39.003 & 2.941 & 2.174 & 0.920 & \underline{38.488} & \underline{4.496} & \underline{3.698} & \underline{0.937} & \underline{43.496} & \underline{0.707} & \underline{0.630} & \underline{0.983} \\  
 \textbf{Proposed} & \textbf{39.345} & \textbf{2.849} & \textbf{2.093} & \textbf{0.922} & \textbf{38.822} & \textbf{4.392} & \textbf{3.556} & \textbf{0.940} & \textbf{43.913} & \textbf{0.685} & \textbf{0.597} & \textbf{0.985} \\ \hline
\end{tabular}
\caption{Comparisons on WV3, QB, and GF2 datasets with 20 reduced-resolution samples, respectively. Best: \textbf{bold}, and second-best: \underline{underline}.}
\label{reduced}
\end{table*}

\subsection{Datasets and Benchmark}

To benchmark the effectiveness of our network for pansharpening, 
we adopt various datasets, 
including datasets captured by the WorldView-3 (WV3), GaoFen-2 (GF2), and QuickBird (QB) sensors.
Since ground truth (GT) images are not available, Wald's protocol \cite{wald1997fusion} is applied. Each training dataset consists of PAN, LRMS, and GT image pairs with sizes of 64 × 64, 16 × 16 × 8, and 64 × 64 × 8, respectively. 
We obtain our datasets and data processing methods from the PanCollection repository\footnote{\url{https://github.com/liangjiandeng/PanCollection}} 
\cite{deng2022machine}.
To evaluate the proposed method, several state-of-the-art pansharpening methods are selected, including three traditional methods, MTF-GLP-FS \cite{MRA}, BDSD-PC \cite{CS}, and TV \cite{TV}, and seven deep-learning methods, including PNN, PanNet \cite{PanNet}, DiCNN, FusionNet \cite{FUsionNet}, U2Net \cite{U2Net}, PanMamba \cite{panMamba}, and CANNet \cite{CANNet}.

\subsection{Implementation Details}
We implement our network using the PyTorch framework on an RTX 4090D GPU. The learning rate is set to \( 9 \times 10^{-4} \) and is halved every 90 epochs. The model is trained for 360 epochs with a batch size of 32. The Adam optimizer is employed. Our method's performance is assessed using standard pansharpening metrics including SAM \cite{SAM}, ERGAS \cite{ERGAS}, and Q4/Q8 \cite{Q4} for reduced-resolution datasets, and HQNR \cite{HQNR}, D$_s$, and D$_\lambda$ for full-resolution datasets.

\subsection{Comparison with State-of-the-Art Methods}

\begin{figure*}[!h]
\centering
\includegraphics[width=0.99\textwidth]{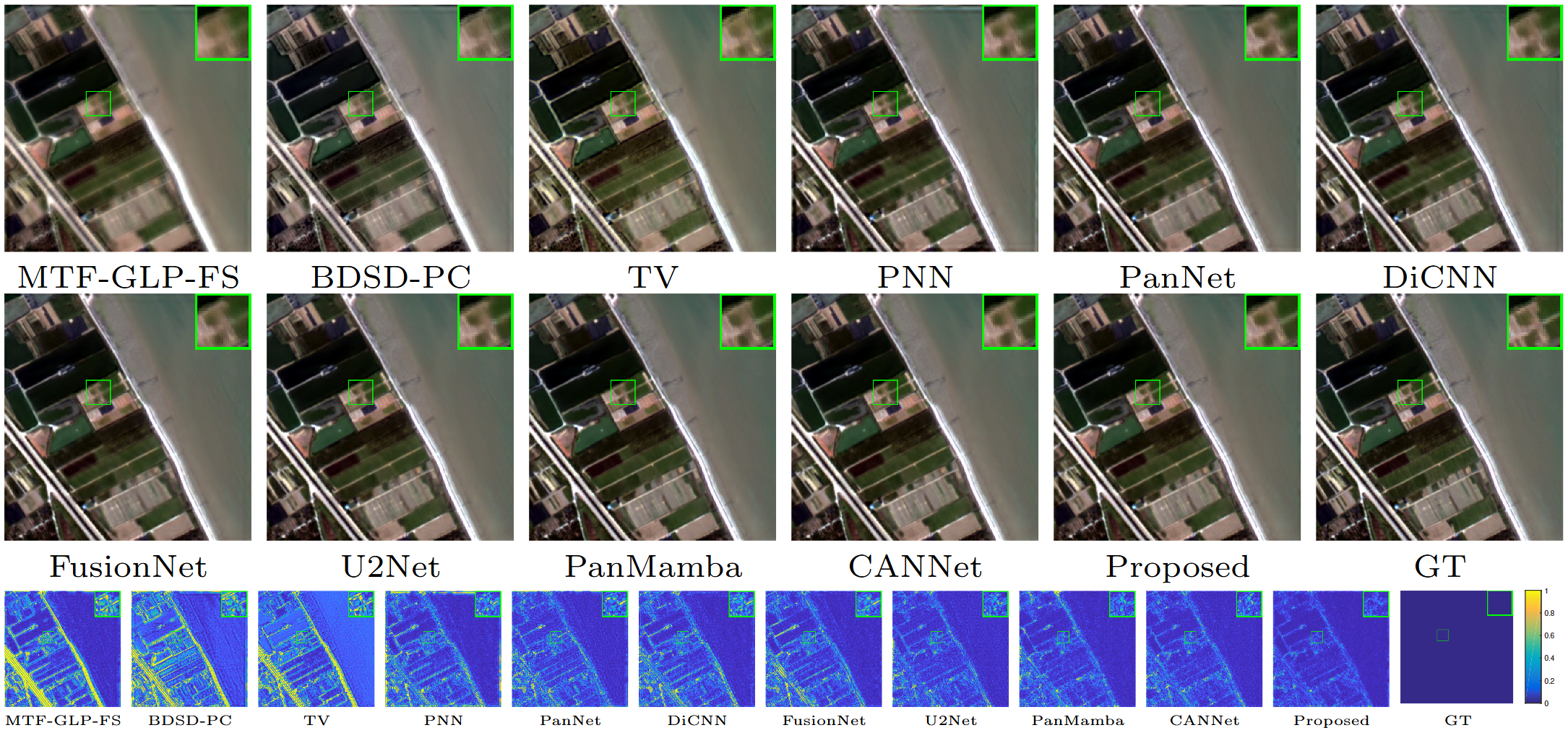} 
\caption{The visual results (Top) and residuals (Bottom) of all compared approaches on the GF2 reduced-resolution dataset.} 
\label{qualitative-gf2}
\end{figure*}

\begin{table*}[h]
\centering
\begin{tabular}{ c| c@{\hskip 0.05in}c@{\hskip 0.05in}c@{\hskip 0.05in}c@{\hskip 0.05in}c@{\hskip 0.05in}c@{\hskip 0.05in}c@{\hskip 0.05in}c@{\hskip 0.05in}c@{\hskip 0.05in}c@{\hskip 0.05in}c }
\hline
\textbf{Metric} & \textbf{MTF-GLP-FS} & \textbf{BDSD-PC} & \textbf{TV} & \textbf{PNN} & \textbf{PanNet} & \textbf{DiCNN} & \textbf{FusionNet} & \textbf{U2Net} & \textbf{PanMamba} & \textbf{CANNet} & \textbf{Proposed} \\ \hline
\textbf{D$_\lambda \downarrow$} & 0.020 & 0.063 & 0.023 & 0.021 & \textbf{0.017} & 0.036 & 0.024 & 0.020 & 0.018 & 0.020 & \textbf{0.017} \\ 
\textbf{D$_s \downarrow$} & 0.063 & 0.073 & 0.039 & 0.043 & 0.047 & 0.046 & 0.036 & \underline{0.028} & 0.053 & 0.030 & \textbf{0.027} \\ 
\textbf{HQNR$\uparrow$} & 0.919 & 0.870 & 0.938 & 0.937 & 0.937 & 0.920 & 0.941 & \underline{0.952} & 0.930 & 0.951  &  \textbf{0.957} \\ \hline

\end{tabular}
\caption{Quantitative comparisons on the WV3 full-resolution dataset.}
\label{WV3_full}
\end{table*}

\subsubsection{Reduced-Resolution Assessment}
Table \ref{reduced} clearly shows a comparison of our proposed method with the current best methods across three datasets. Our method
consistently achieves the best results across all metrics. Specifically, our method achieves a PSNR improvement of 0.228dB, 0.334dB, and 0.417dB on the WV3, QB, and GF2 datasets, respectively, compared to the second-best results. These improvements highlight the clear advantages of our method, confirming its competitiveness in the field. Fig.~\ref{qualitative-wv3} and Fig.~\ref{qualitative-gf2} show the qualitative assessment results for two datasets and their corresponding ground truth (GT). By comparing the MSE residuals between the pan-sharpened results and the ground truth, it is evident that our residual maps are the darkest, indicating that our method outperforms others. The experimental results above demonstrate that our method is superior to the latest state-of-the-art pansharpening methods.

\begin{table}[H]
\centering
\setlength{\tabcolsep}{4pt}
\begin{tabular}{c c c c c}
\hline
\textbf{Ablation} & \textbf{PSNR$\uparrow$} & \textbf{SAM$\downarrow$} & \textbf{ERGAS$\downarrow$} & \textbf{Q8$\uparrow$} \\ \hline
Frequency-Query & 38.884 & 2.968 & 2.206 & 0.918 \\ 
Spatial-Key & 39.156 & 2.901 & 2.138 & 0.921 \\ 
Fusion-Value & 38.921 & 2.971 & 2.203 & 0.918 \\ 
\textbf{Ours} & \textbf{39.345} & \textbf{2.849} & \textbf{2.093} & \textbf{0.922} \\ 
\hline
\end{tabular}
\caption{Ablation experiment about Attention Triplet on WV3 reduced-resolution dataset.}
\label{sao}
\end{table}

\subsubsection{Full-Resolution Assessment}
To demonstrate the generalization ability of our method, we conduct experiments on full-resolution samples of WV3. The quantitative evaluation results are shown in Table \ref{WV3_full}. Our method achieves the best HQNR results, which reflects its ability to balance spectral and spatial distortions, demonstrating its high application value.

\begin{table}[h]
\centering
\begin{tabular}{ l c c c c}
\hline
\textbf{Ablation} & \textbf{PSNR$\uparrow$} & \textbf{SAM$\downarrow$} & \textbf{ERGAS$\downarrow$} & \textbf{Q8$\uparrow$} \\ \hline
MFFA & 38.486 & 3.148 & 2.349 & 0.915 \\ 
SDEM & 38.993 & 2.937 & 2.178 & 0.918 \\ 
Multi-Scale & 38.851 & 2.995 & 2.214 & 0.916 \\ 
FAB & 39.074 & 2.919 & 2.155 & 0.919 \\ 
 \textbf{Ours} & \textbf{39.345} & \textbf{2.849} & \textbf{2.093} & \textbf{0.922} \\ \hline
\end{tabular}
\caption{Ablation experiment about key components and strategy on WV3 reduced-resolution dataset.}
\label{ablation}
\end{table}

\subsection{Ablation Study}
This section explores the rationale behind the design of the Frequency Attention Triplet and the roles of key components and strategies in WFANet. We conduct a series of ablation experiments on the WV3 dataset to demonstrate their effectiveness and validity. First, Table \ref{sao} presents three sets of ablation experiments for Frequency Attention Triplet, followed by Table \ref{ablation}, which shows four sets of ablation experiments for WFANet. \textit{More experiments and detailed ablation settings can be found in the supplementary materials.}

\subsubsection{Frequency Attention Triplet}
To demonstrate the effectiveness of Frequency-Query, we remove the DWT operation and directly use a spatial domain \(Q\) instead of using \(Q_i\) in the different frequency domains. For Spatial-Key, we no longer use \(P_{LL}\) obtained by DWT, which represents the overall spatial features, but instead use features from \(P\) after convolution, introducing interference from high-frequency information. Regarding Fusion-Value, we no longer use the fusion of LRMS and \(P_{LL}\), but instead include only the information from LRMS, thereby lacking the low-frequency spatial information represented by \(P_{LL}\). The results in Table \ref{sao} demonstrate the effectiveness of each component in the Frequency Attention Triplet.

\subsubsection{Multi-Frequency Fusion Attention}
To demonstrate the effectiveness of the attention mechanism in MFFA, we replace it with the convolutional network, where HRMS and different frequency features are concatenated and then processed through convolution. The results in Table \ref{ablation} prove it.

\subsubsection{Spatial Detail Enhancement Module}
The SDEM enhances reconstructed images by injecting spatial details from frequency domains. We remove the SDEM while retaining the MFFA, and the lack of spatial detail is noticeable. The results in Table \ref{ablation} confirm the effectiveness of the SDEM.

\subsubsection{The Multi-Scale Training Strategy}
To demonstrate the effectiveness of the multi-scale strategy, we replace the multi-scale network in this paper with a single-scale network, aligning the sizes using convolution operations. The results in Table \ref{ablation} confirm the significance of this strategy.

\subsubsection{Frequency Adaptation Block}
We explore how operations adapt to frequency domains. Fig.~\ref{comparison} shows two network architectures for the SDEM. We replace the FABs with Convolution Blocks, and Table \ref{ablation} shows the superior feature extraction across frequencies provided by the FABs.

\section{Conclusion}
In this paper, we propose a novel approach, Multi-Frequency Fusion Attention (MFFA), which leverages an effective method for frequency decomposition and reconstruction. By designing Frequency-Query, Spatial-Key, and Fusion-Value with clear physical significance, MFFA achieves more effective and precise fusion in the frequency domain. We also emphasize the adaptation of different operations to the frequency domain and have designed a comprehensive multi-scale fusion strategy. Ablation experiments further confirm the effectiveness of our approach. Extensive experiments on three different satellite datasets demonstrate that our model outperforms state-of-the-art methods.

\section{Acknowledgments}
This research is supported by the National Natural Science Foundation of China (12271083), and the Natural Science Foundation of Sichuan Province (2024NSFSC0038).

\bibliography{aaai25}





\clearpage
\twocolumn[\begin{minipage}{.99\textwidth}\centering%
    \LARGE\bf Wavelet-Assisted Multi-Frequency Attention Network for Pansharpening     Supplemental Material\vspace*{2ex}

\end{minipage}]

\maketitle
\begin{abstract}
 
The supplementary materials provide additional insights into the method proposed in our paper.
First, we offer a more detailed explanation of the discrete wavelet transform (DWT) and multi-scale strategy employed in this work.
Furthermore, we present additional experiments and discussions, including a comparison of parameter numbers and further validation of the Frequency Attention Triplet.
Lastly, we provide a more comprehensive overview of the ablation experiment settings and offer additional quantitative and qualitative comparisons of different methods.

\end{abstract}

\section{Method Supplementary}

\subsection{Detailed Explanation of Wavelet Transform}
There are various forms of wavelet transform\cite{wave1,wave2,wave3}, and in our method, we utilize the discrete wavelet transform (DWT). This process can be illustrated with a simple example.
Consider an image, \( I \), represented by the following matrix:
\begin{equation}
I =
\begin{bmatrix}
a_{11} & a_{12} \\
a_{21} & a_{22} 
\end{bmatrix}
\end{equation}
Upon applying the DWT, the resulting low-frequency component can be expressed as:
\begin{equation}
LL = \frac{a_{11} + a_{12} + a_{21} + a_{22}}{4}
\end{equation}
The high-frequency components in the horizontal, vertical, and diagonal directions are represented by the LH, HL, and HH components, respectively, and are computed as follows:
\begin{equation}
LH = \frac{a_{11} + a_{12} - a_{21} - a_{22}}{4}
\end{equation}

\begin{equation}
HL = \frac{a_{11} - a_{12} + a_{21} - a_{22}}{4}
\end{equation}

\begin{equation}
HH = \frac{a_{21} + a_{22} - a_{11} - a_{12}}{4}
\end{equation}
\begin{figure}[!h]
\centering
\includegraphics[width=\columnwidth]{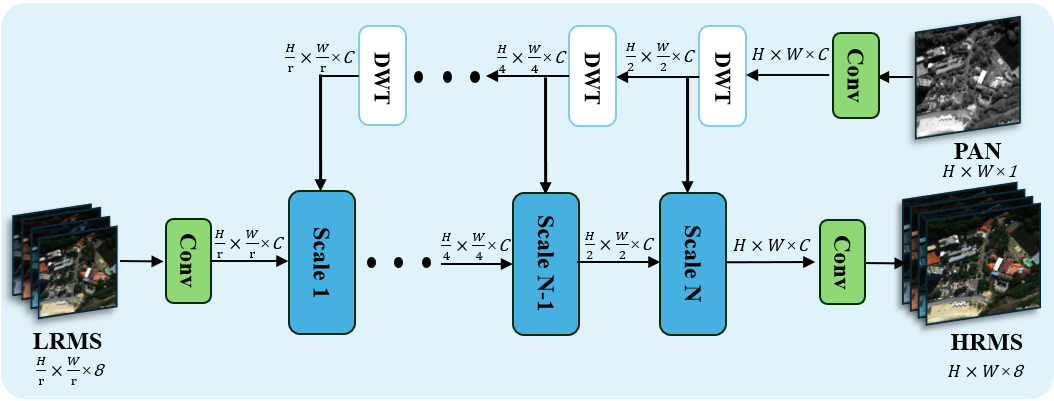} 
\caption{The explanation of the multi-scale processing adopted by our method. The dimensions above the arrows represent the data dimensions entering the module to which the arrow points. Each scale differs by a factor of 2, so \(r\) is a multiple of 2. In this paper, \(C = 32\).}
\label{scale}
\end{figure}
For larger matrices, the image is divided into multiple 2x2 regions, and each region is processed individually using the above method. 
The inverse discrete wavelet transform (IDWT) reconstructs the original image by combining the LL, LH, HL, and HH components through a system of equations, ensuring the accurate and lossless recovery of the image.
To create a wavelet pyramid, the DWT operation is applied recursively to the LL component from the previous scale, constructing the pyramid through multiple iterations.
In practice, these operations can be efficiently implemented using fixed-value convolution and deconvolution kernels.

\subsection{Multi-Scale Strategy Details}
Previous works \cite{8281501,jin2022laplacian} have explored the advantages of leveraging this multi-scale property.
The multi-scale characteristic of the wavelet pyramid allows for more effective utilization of features across different scales. The multi-scale fusion process we adopt is illustrated in Fig.~\ref{scale}. Using multiple DWT operations, we obtain PAN image features with gradually decreasing scales until they match the scale of the LRMS. Fusion begins at the smallest scale and progressively transitions to the largest scale. At each scale, the input features have the same dimensions, and the output dimensions are twice those of the input, thereby achieving a progressive fusion process.

\begin{table*}[h]
\centering
\begin{tabular}{ c|c c c c c c c c c }
\hline
\textbf{Methods} & \textbf{PNN} & \textbf{PanNet} & \textbf{DiCNN} & \textbf{FusionNet} & \textbf{U2Net} & \textbf{PanMamba} & \textbf{CANNet} & \textbf{WFANet} & \textbf{WFANet-L}  \\ \hline
\textbf{Params(M)} & 0.10 & 0.08 & 0.04 & 0.08 & 0.66 & 0.48 & 0.78 & 0.52 & 0.07 \\ \hline
\end{tabular}
\caption{Comparison of parameters for different methods.}
\label{params}
\end{table*}

\section{Experimental Supplementary}

\subsection{ Comparison of Parameter Numbers
}
In this section, we compare the parameter numbers of various DL-based pansharpening methods, as illustrated in Table \ref{params}.
We divide DL-based pansharpening methods into two categories based on their number of parameters. Models with no more than 0.10M parameters are designated as lightweight networks,
 whereas those exceeding 0.10M parameters are classified as heavyweight networks. WFANet belongs to the heavyweight category, and we also designed a lightweight version, WFANet-L.
To reduce network parameters while maintaining performance, we decreased the common channel size from 32 to 24 and simplified several MLP layers. 
To ensure a fair comparison, Fig.~\ref{comparison} shows the results of the lightweight networks in the left half and the heavyweight networks in the right half, with PSNR representing the model performance\cite{8281501}.
 Both WFANet-L and WFANet achieve strong performance while maintaining a relatively low number of parameters. These results demonstrates that our method effectively balances model performance with manageable complexity.

\subsection{Further Validation of the Frequency Attention Triplet}
To further validate the effectiveness of the Frequency Attention Triplet design, we conducted experiments where we systematically swapped the roles of Frequency-Query, Spatial-Key, and Fusion-Value as the Query, Key, and Value in the attention mechanism. This resulted in six different configurations. The original configuration is labeled as Ours, while the alternative configurations, named V1 through V5, \textit{each represents a specific permutation of Frequency-Query, Spatial-Key, and Fusion-Value serving as Query, Key, and Value, respectively}. As shown in Table \ref{qkv}, the experimental results clearly demonstrate that our method achieves the best performance across all metrics, which can be attributed to the thoughtful design based on their physical significance.

\begin{figure}[t]
\centering
\includegraphics[width=0.99\columnwidth]{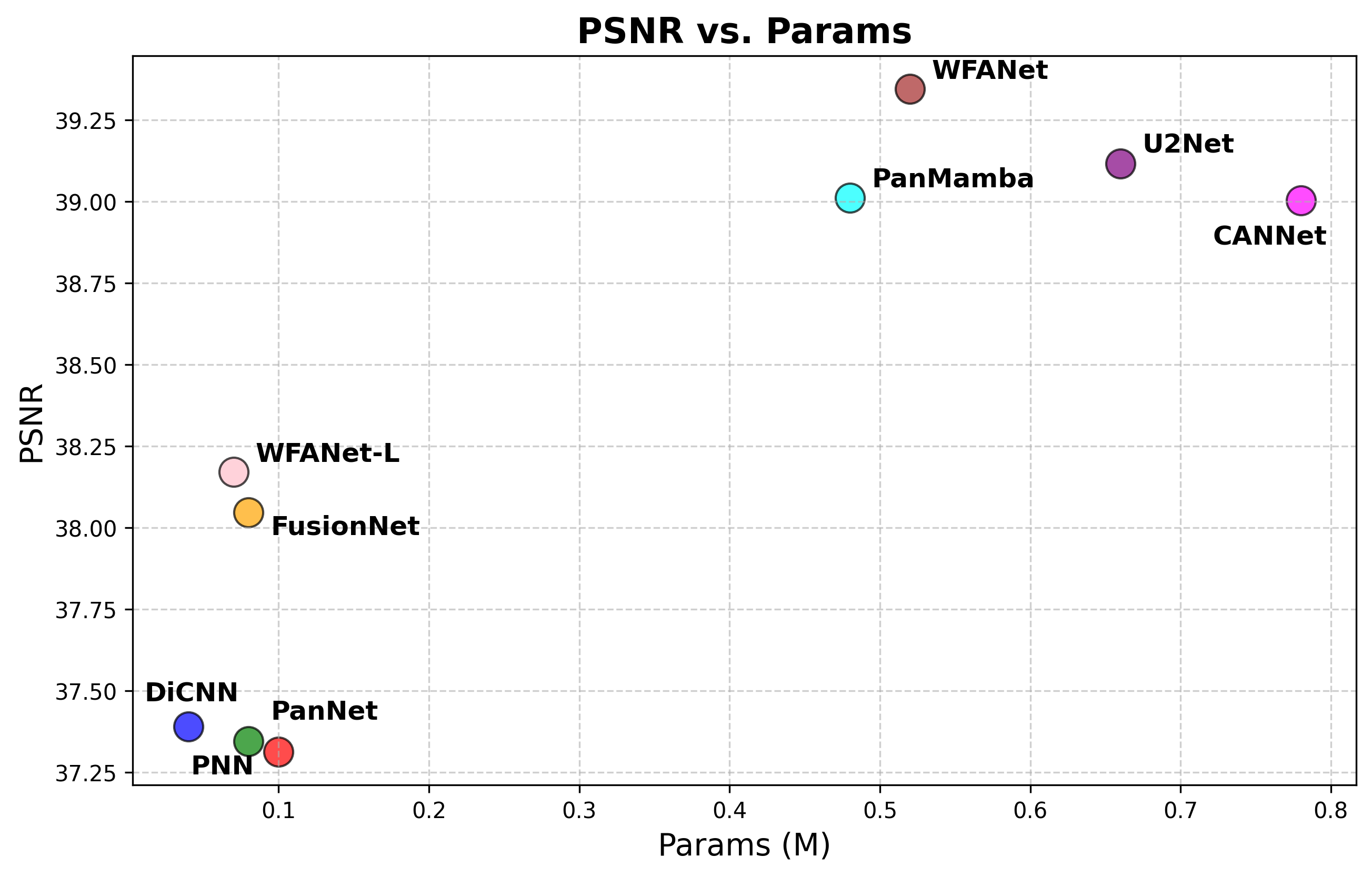} 
\caption{Trade-off between parameter numbers and PSNR on the WV3 reduced-resolution dataset. The left side shows lightweight networks, while the right side shows heavyweight networks.
} 
\label{comparison}
\end{figure}

\begin{table}[t]
\centering
\begin{tabular}{ c c c c c }
\hline
\textbf{Method} & \textbf{PSNR$\uparrow$} & \textbf{SAM$\downarrow$} & \textbf{ERGAS$\downarrow$} & \textbf{Q8$\uparrow$} \\ \hline
V1 & 39.092 & 2.925 & 2.161 & 0.920 \\ 
V2 & 38.954 & 2.946 & 2.190 & 0.917 \\ 
V3 & 38.916 & 2.954 & 2.198 & 0.918 \\ 
V4 & 39.056 & 2.914 & 2.163 & 0.920 \\ 
V5 & 38.573 & 3.104 & 2.316 & 0.915 \\ 
\textbf{Ours} & \textbf{39.345} & \textbf{2.849} & \textbf{2.093} & \textbf{0.922} \\ \hline
\end{tabular}
\caption{Comparison of different methods for further validation of the Frequency Attention Triplet.}
\label{qkv}
\end{table}

\begin{figure}[t]
\centering
\includegraphics[width=0.99\columnwidth]{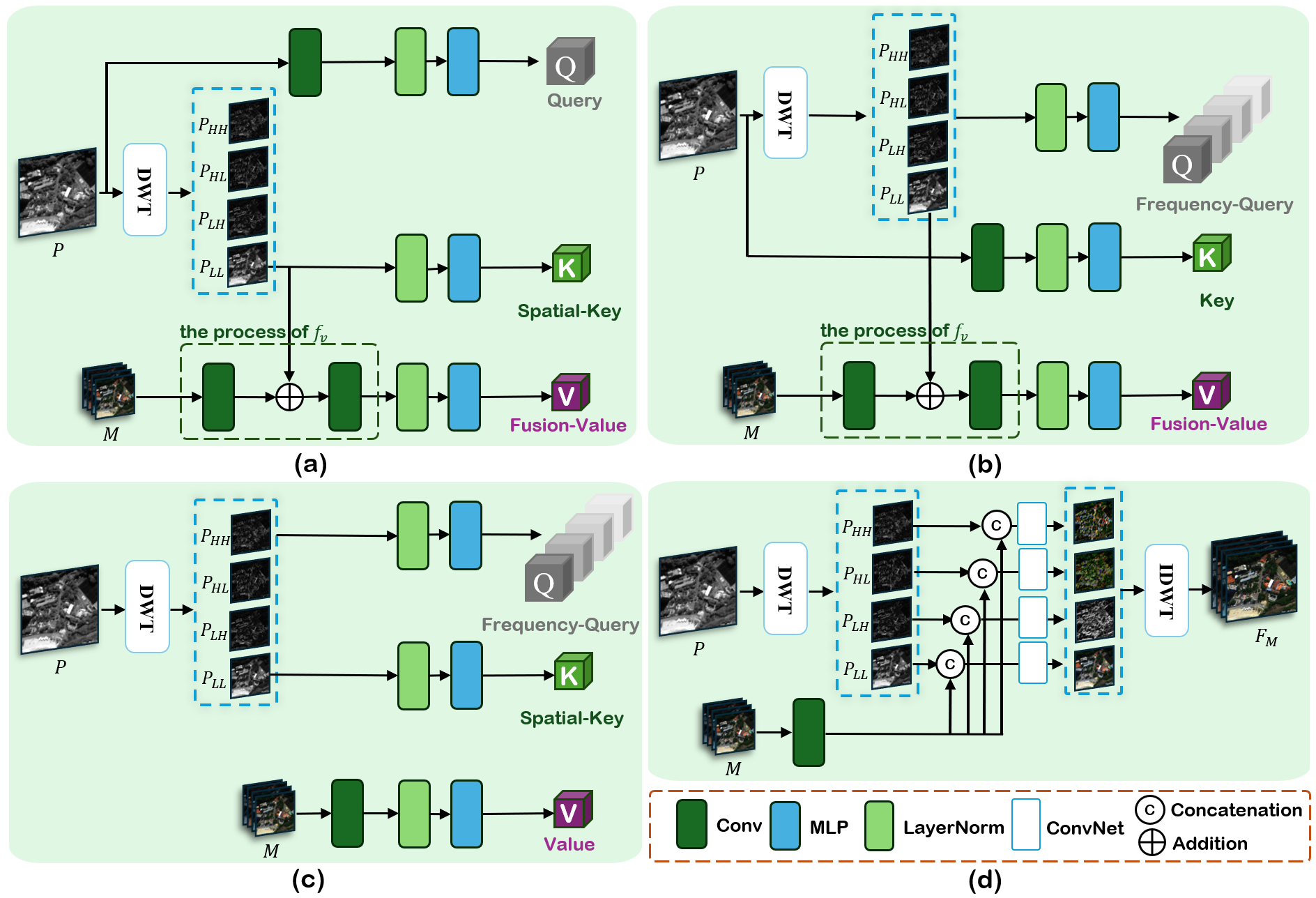} 
\caption{A detailed explanation of the ablation experiments: (a) Frequency-Query ablation, (b) Spatial-Key ablation, (c) Fusion-Value ablation, and (d) MFFA ablation.} 
\label{ablation}
\end{figure}

\begin{table*}[t]
\centering
\begin{tabular}{ c| c@{\hskip 0.05in}c@{\hskip 0.05in}c@{\hskip 0.05in}c@{\hskip 0.05in}c@{\hskip 0.05in}c@{\hskip 0.05in}c@{\hskip 0.05in}c@{\hskip 0.05in}c@{\hskip 0.05in}c@{\hskip 0.05in}c }
\hline
\textbf{Metric} & \textbf{MTF-GLP-FS} & \textbf{BDSD-PC} & \textbf{TV} & \textbf{PNN} & \textbf{PanNet} & \textbf{DiCNN} & \textbf{FusionNet} & \textbf{U2Net} & \textbf{PanMamba} & \textbf{CANNet} & \textbf{Proposed} \\ \hline
\textbf{D$_\lambda \downarrow$} & 0.035 & 0.076 & 0.055 & 0.032 & \underline{0.018} & 0.037 & 0.035 & 0.024 & 0.023 & 0.019 & \textbf{0.017} \\ 
\textbf{D$_s \downarrow$} & 0.143 & 0.155 & 0.112 & 0.094 & 0.080 & 0.099 & 0.101 & \textbf{0.051} & 0.057 & 0.063 & \underline{0.052} \\ 
\textbf{HQNR$\uparrow$} & 0.828 & 0.781 & 0.839 & 0.877 & 0.904 & 0.868 & 0.867 & \underline{0.927} & 0.921 & 0.919 & \textbf{0.932} \\ \hline

\end{tabular}
\caption{Quantitative comparisons on the GF2 full-resolution dataset.}
\label{WV3_full}
\end{table*}

\begin{figure*}[!h]
\centering
\includegraphics[width=0.99\textwidth]{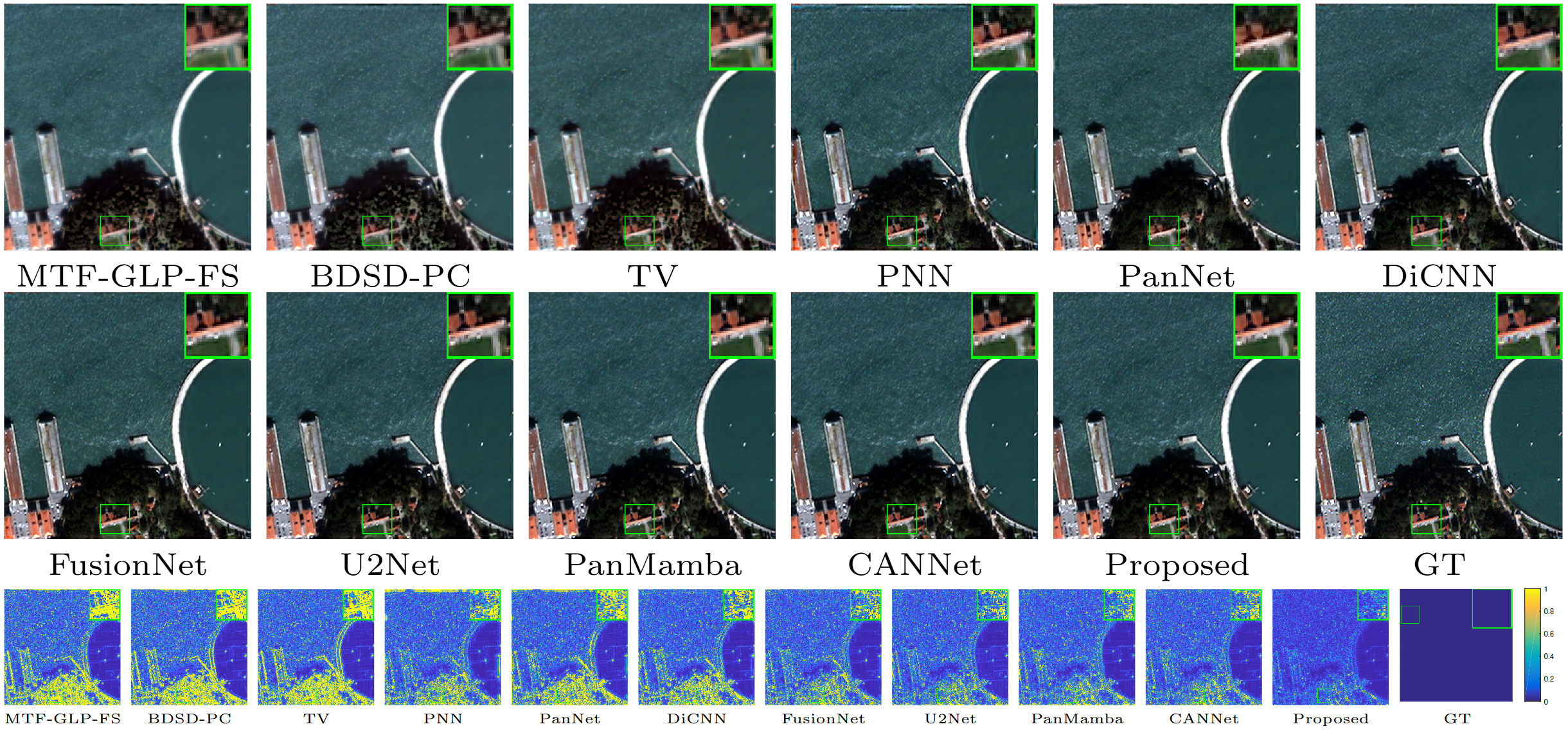} 
\caption{The visual results (Top) and residuals (Bottom) of all compared approaches on the QB reduced-resolution dataset.} 
\label{reduce-wv3}
\end{figure*}

\begin{figure*}[!h]
\centering
\includegraphics[width=0.99\textwidth]{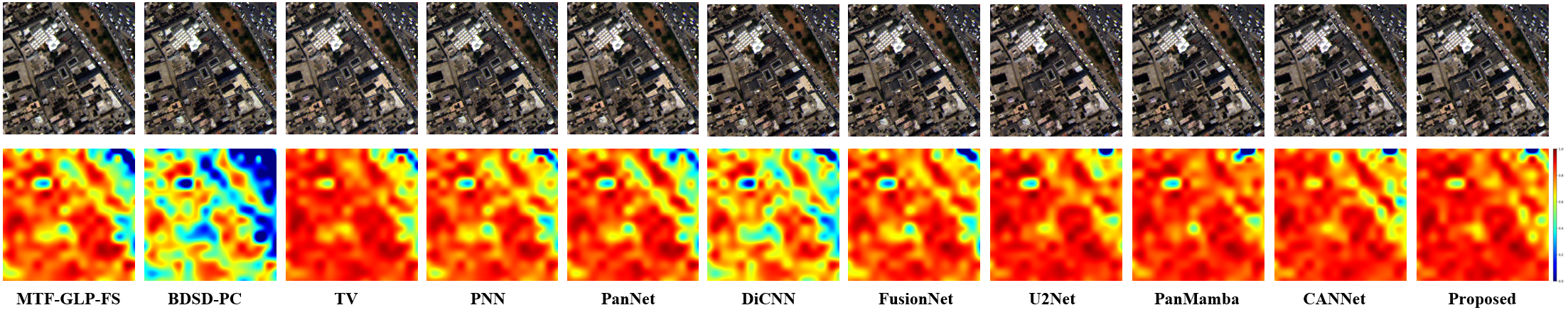} 
\caption{The visual results (Top) and  HQNR maps (Bottom) of all compared approaches on the WV3 full-resolution dataset. } 
\label{full-wv3}
\end{figure*}

\begin{figure*}[!h]
\centering
\includegraphics[width=0.99\textwidth]{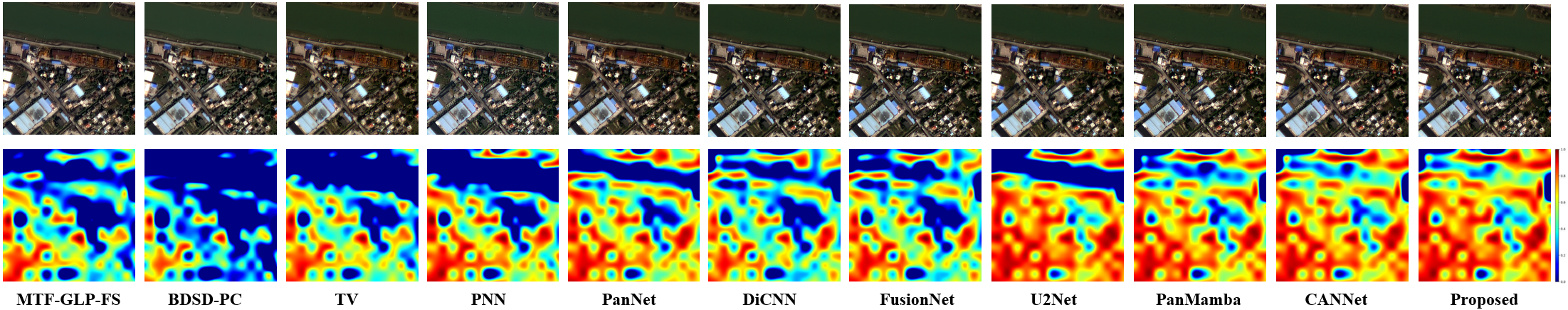} 
\caption{The visual results (Top) and  HQNR maps (Bottom) of all compared approaches on the GF2 full-resolution dataset. } 
\label{full-GF2}
\end{figure*}

\subsection{Ablation Settings Details}
This section provides a more detailed description of some ablation experiments.
\subsubsection{Frequency Attention Triplet}
First, without any ablation, the generation process of our Frequency Attention Triplet is as follows:
\begin{equation}
\begin{aligned}
    &Q_i = \operatorname{MLP}(\operatorname{LN}(P_i)) \\
    &K = \operatorname{MLP}(\operatorname{LN}(P_{LL})) \\
    &V = \operatorname{MLP}(\operatorname{LN}(f_v(M, P_{LL})))
\end{aligned}
\end{equation}
\textit{We separately altered the generation method of one component within the Frequency Attention Triplet while keeping the others unchanged}. Fig.~\ref{comparison} (a)-(c) correspond to the three different ablation settings, where each substitutes \( \overline{Q} \), \( \overline{K} \), or \( \overline{V} \) for the original component while keeping the other two components unchanged.
The process can be obtained using the following equations:
\begin{equation}
\begin{aligned}
    &\overline{Q} = \operatorname{MLP}(\operatorname{LN}(\operatorname{Conv}(P))) \\
    &\overline{K} = \operatorname{MLP}(\operatorname{LN}(\operatorname{Conv}(P))) \\
    &\overline{V} = \operatorname{MLP}(\operatorname{LN}(\operatorname{Conv}(M)))
\end{aligned}
\end{equation}

\subsubsection{Multi-Frequency Fusion Attention}
As illustrated in Fig.~\ref{comparison} (d), we respectively concatenate the convolved \(M\) with the features in different frequency domains and then extract features through a convolutional network. The design of this convolutional network follows the classical PNN approach \cite{PNN}.

\subsection{Additional Results}
In this section, we present additional qualitative and quantitative results. 
Table \ref{WV3_full} presents the results on the GF2 full-resolution dataset. Fig.~\ref{reduce-wv3} illustrates the visualization results on the QB reduced dataset. Fig.~\ref{full-wv3} and Fig.~\ref{full-GF2} present the visualization results of the WV3 and GF2 full-resolution datasets. 
As depicted in the second row, the redder areas indicate better performance, while the bluer areas indicate poorer performance. Among the methods compared, ours shows the largest and deepest red area, indicating the best performance.

\clearpage

\end{document}